\documentclass[showpacs,twocolumn,preprintnumbers,amsmath,amssymb,superscriptaddress,nofootinbib]{revtex4}
\usepackage{graphicx}
\usepackage{bm}
\usepackage{ulem}
\usepackage{color}

\newcommand{\simgt}{\lower.5ex\hbox{$\; \buildrel > \over \sim \;$}}
\newcommand{\simlt}{\lower.5ex\hbox{$\; \buildrel < \over \sim \;$}}

\newcommand{\morv}[1]{#1}

\begin{document}

\title{Understanding caustic crossings in giant arcs: characteristic
  scales, event rates, and constraints on compact dark matter}

\author{Masamune Oguri}
\affiliation{
Research Center for the Early Universe, University of Tokyo, Tokyo 113-0033, Japan}
\affiliation{
Department of Physics, University of Tokyo, Tokyo 113-0033, Japan}
\affiliation{
Kavli Institute for the Physics and Mathematics of the Universe (Kavli
IPMU, WPI), The University of Tokyo, Chiba 277-8582, Japan} 
\author{Jose M. Diego}
\affiliation{
IFCA, Instituto de F\'{i}sica de Cantabria (UC-CSIC), Av. de
Los Castros s/n, 39005 Santander, Spain}
\author{Nick Kaiser}
\affiliation{
Institute for Astronomy, University of Hawaii, 2680 Woodlawn Drive,
Honolulu, HI 96822-1839, USA} 
\author{Patrick L. Kelly}
\affiliation{
School of Physics and Astronomy, University of Minnesota, 116 Church Street SE, Minneapolis, MN 55455, USA}
\author{Tom Broadhurst}
\affiliation{
Department of Theoretical Physics, University of the Basque Country, Bilbao 48080, Spain}
\affiliation{
IKERBASQUE, Basque Foundation for Science, Alameda Urquijo, 36-5 48008 Bilbao, Spain}

\date{\today}

\begin{abstract}
The recent discovery of fast transient events near critical curves of
massive galaxy clusters, which are interpreted as highly magnified
individual stars in giant arcs due to caustic crossing, opens up the
possibility of using such microlensing events to constrain a range of
dark matter models such as primordial black holes and scalar field
dark matter. Based on a simple analytic model, we study lensing
properties of a point mass lens embedded in a high magnification
region, and derive the dependence of the peak brightness, microlensing
time scales, and event rates on the mass of the point mass lens as
well as the radius of a source star that is magnified. We find that 
the lens mass and source radius of the first event MACS~J1149 Lensed
Star 1 (LS1) are constrained, with the lens mass range of
$0.1~M_\odot \lesssim M \lesssim 4\times 10^3M_\odot$  and the source radius
range of $40~R_\odot \lesssim R \lesssim 260~R_\odot$. In the most
plausible case with $M\approx 0.3~M_\odot$ and $R\approx 180~R_\odot$,
the source star should have been magnified by a factor of 
$\approx 4300$ at the peak. The derived lens properties are fully
consistent with the interpretation that MACS~J1149 LS1 is a
microlensing event produced by a star that contributes to the
intra-cluster light. We argue that compact dark matter models with
high fractional mass densities for the mass range $10^{-5}M_\odot
\lesssim M\lesssim 10^2M_\odot$ are inconsistent with the observation of
MACS~J1149 LS1 because such models predict too low magnifications.
Our work demonstrates a potential use of caustic 
crossing events in giant arcs to constrain compact dark matter.
\end{abstract}


\maketitle

\section{Introduction}
\label{sec:intro}

Recently, Kelly et al. \cite{Kelly:2017fps} reported the discovery of
MACS~J1149 Lensed Star 1 (LS1, also known as ``Icarus''), a faint
transient near the critical curve of the massive cluster
MACS~J1149.6+2223. The transient is interpreted as a luminous star in
the host galaxy of supernova Refsdal \cite{Kelly:2014mwa} at $z=1.49$,
which is magnified by a compact object very close to the critical
curve of the foreground lens. The light curve is consistent with caustic
crossing of the background star, and from the comparison with
ray-tracing simulations it was suggested that the star was probably
magnified by a factor of several thousands at the peak
brightness. There was also an additional transient (``Iapyx'') detected at
roughly the same distance from the critical curve of the cluster but
on the opposite side, which can be the counterimage of LS1. Furthermore,
Rodney et al. \cite{Rodney:2017aaa} reported two peculiar fast
transients (``Spock'') behind the cluster MACS~J0416.1$-$2403 in a
strongly lensed galaxy at $z=1.0054$. While the Spock events can be
explained by the outburst of a Luminous Blue Variable star or a
recurrent nova, one possible interpretation is that these two events
are also caustic crossing events.   

These caustic crossing events in giant arcs behind massive clusters
remarkably differ from traditional microlensing
observations. Microlensing observations in our Galaxy or in
nearby galaxies (e.g.,
\cite{Alcock:1993qc,Udalski:1994ei,Alcock:2000ph,Sumi:2002wg,deJong:2005jm,Tisserand:2006zx,Griest:2013esa,Niikura:2017zjd})
are usually produced by isolated stars (or compact objects), whereas
caustic crossing events in giant arcs are produced by stars embedded
in high magnification regions due to the cluster potential. As shown
in previous work (e.g.,  
\cite{Chang:1979zz,Paczynski:1986aa,Kayser:1986aa,Witt:1990aa,Schechter:2002dm}),
microlensing properties are significantly modified by the presence of
such convergence and shear field due to the cluster potential.

Perhaps quasar microlensing (e.g.,
\cite{Wambsganss:1990aa,Schmidt:1998ah,Pooley:2006rh,Chartas:2008cg,Morgan:2010xf,Blackburne:2010ja,Jimenez-Vicente:2015eva,Mediavilla:2017bok})
more resembles caustic crossing events in giant arcs in the sense that
it is also caused by stars embedded in high magnification regions due
to the lensing galaxy. However there are several notable differences
between microlensing in giant arcs and quasar microlensing. For
example, in the former case the source is a star, whereas in the
latter case the source is a quasar. While the internal structure of a
quasar is complicated with largely different sizes for different
wavelengths, the surface brightness distribution of a star is uniform
(neglecting limb darkening) and its radius can be predicted from
observations of nearby stars. Also 
the radius of a star is typically much smaller than the size of the light
emitting region of a quasar. The smaller radius translates into the
higher maximum magnification that scales as the inverse of the square
root of the radius (see Section~\ref{sec:general}). In addition, while
typical magnifications of the brightest images of lensed quasars are
$\approx 10-20$, MACS~J1149 LS1 is observed very close to the
critical curve of the macro lens model where the magnification due to
the macro lens model is estimated to be $\gtrsim 300$. Such high
magnification of the macro mass model is possible because the giant
arc directly crosses the critical curve and many stars in the giant
arcs are located very near the critical curve.

Caustic crossing events in giant arcs are new phenomena that may probe
a different parameter space from previously known microlensing events.
This possibility of observing highly magnified images of individual
stars in giant arcs was first discussed in \cite{MiraldaEscude:1991aa}, 
although only the smooth cluster potential was considered in that
work. Motivated by the discovery of MACS~J1149 LS1,
\cite{Diego:2017drh} and \cite{Venumadhav:2017pps} revisited this
problem, and argued that even a small fraction of compact objects in
the lens disrupts the critical curve into a network of micro-caustics,
and drastically modifies the microlensing properties near the critical
curve. Even if dark matter consists entirely of a smooth component,
such compact objects are expected to exist, like for instance the
stars that are responsible for the so-called intra-cluster light
(ICL). Given the drastic change of 
lensing properties near the critical curve, it has been argued that
caustic crossing events in giant arcs may serve as a powerful probe of
a range of dark matter scenarios such as Primordial Black Holes (PBHs)
\cite{Bird:2016dcv,Sasaki:2016jop,Kawasaki:2016pql,Carr:2016drx,Inomata:2017okj}
and scalar field dark matter \cite{Hu:2000ke,Schive:2014dra,Hui:2016ltb}. 

In this paper, we adopt a simple analytic model that consists of
a point mass lens and a constant convergence and shear component, and
study basic microlensing properties of this lens model. The result is
used to derive characteristic scales of caustic crossing events in
giant arcs, and their dependences on the lens and source properties. 
The result is used to interpret MACS~J1149 LS1 and place constraints
on the lens mass as well as the property of the source star. We also
discuss the event rates of caustic crossing. Such analytic studies of
caustic crossing events should complement ray-tracing simulations
presented in \cite{Kelly:2017fps} and \cite{Diego:2017drh} for which
repeating simulations with many different model parameters may be
computationally expensive. Our approach more resembles that in
\cite{Venumadhav:2017pps}, which appeared when this work was almost
completed. 

This paper is organized as follows. We present an analytic lens model
which sets the theoretical background in Section~\ref{sec:general}.
We then discuss what kind of constraints we can place from observed
caustic crossing events in Section~\ref{sec:expected}.
We apply our results to MACS~J1149 LS1 to derive constraints on lens
and source properties of this particular microlensing event in
Section~\ref{sec:constraint}. We discuss event rates and derive
expected event rates for MACS~J1149 LS1 as well as for general cases
in Section~\ref{sec:eventrate}. In Section~\ref{sec:compactdm}, 
we discuss constraints on compact dark matter in the presence of ICL. 
Finally we summarize our results in Section~\ref{sec:summary}. 
Throughout the paper we adopt a cosmological model with the matter
density $\Omega_{\rm m}=0.3$, cosmological constant $\Omega_\Lambda=0.7$, 
and the Hubble constant $H_0=70~{\rm km\,s^{-1}Mpc^{-1}}$. 

\section{General theory}
\label{sec:general}

Here we summarize basic properties of gravitational lensing by a point
mass lens embedded in a high magnification region. The high
magnification region concentrates around the caustics that are
produced by a macro lens model of e.g., a massive cluster of galaxies,
and we consider a perturbation by a compact object to a highly
magnified background object (star) near the caustics. The lensing
properties of such compound system have been studied at depth in the
literature
\cite{Chang:1979zz,Paczynski:1986aa,Kayser:1986aa,Witt:1990aa,Schechter:2002dm},
and more recently by \cite{Diego:2017drh,Venumadhav:2017pps} in the
context of interpreting MACS~J1149 LS1 \cite{Kelly:2017fps}. We
present some key results which are necessary for the discussions in
the following sections. 

\subsection{Lens equation}

We consider a point mass lens in a constant convergence
($\bar{\kappa}$) and shear ($\bar{\gamma}$) field which comes from a
macro lens model. Then we have
\begin{eqnarray}
\mu_{\rm t}^{-1}&=& 1-\bar{\kappa}-\bar{\gamma},\\
\mu_{\rm r}^{-1}&=& 1-\bar{\kappa}+\bar{\gamma}.
\end{eqnarray}
The total magnification by the macro lens model is $\bar{\mu}=\mu_{\rm
  t}\mu_{\rm r}$. We consider a region near the tangential critical
curve where $\mu_{\rm t}^{-1}\approx 0$ gives rise to the high
magnification. 

A point mass lens with mass $M$, in absence of the macro mass model,
has the Einstein radius of 
\begin{equation}
  \theta_{\rm E}=\left(\frac{4GM}{c^2}\frac{D_{ls}}{D_{os}D_{ol}}\right)^{1/2}.
\label{eq:einrad}
\end{equation}
The lens equation for the point mass lens embedded in the macro lens
model is 
\begin{eqnarray}
  \beta_1 & = &
  \frac{\theta_1}{\mu_{\rm r}}-\frac{\theta_{\rm E}^2\theta_1}{\theta^2},\label{eq:beta1}\\
  \beta_2 & = & \frac{\theta_2}{\mu_{\rm t}}-\frac{\theta_{\rm E}^2\theta_2}{\theta^2}.\label{eq:beta2}
\end{eqnarray}
Note that ($\beta_1$, $\beta_2$) is the position of the source and
($\theta_1$, $\theta_2$) is the position of the image. The origin of
the coordinates is taken at the position of the point mass lens. For
simplicity we assume that the shear direction is aligned with the $x$-axis. The
inverse magnification matrix is 
\begin{equation}
  \frac{\partial \vec{\beta}}{\partial \vec{\theta}}
=\left(
    \begin{array}{cc}
     \mu_{\rm r}^{-1}+\theta_{\rm E}^2\cos (2\phi)/\theta^2
    & -\theta_{\rm E}^2\sin (2\phi)/\theta^2 \\
      -\theta_{\rm E}^2\sin (2\phi)/\theta^2
    & \mu_{\rm t}^{-1} -\theta_{\rm E}^2\cos (2\phi)/\theta^2
    \end{array}
  \right),
\end{equation}
where $\phi$ is the polar angle of ($\theta_1$, $\theta_2$).
The magnification is
\begin{equation}
\mu^{-1}=(\mu_{\rm t}\mu_{\rm r})^{-1}-(\mu_{\rm r}^{-1}-\mu_{\rm
  t}^{-1})\cos(2\phi)\frac{\theta_{\rm E}^2}{\theta^2}-\frac{\theta_{\rm E}^4}{\theta^4}.
\end{equation}

\subsection{Critical curve and caustic}

\begin{figure*}[t]
\begin{center}
\includegraphics[width=0.8\textwidth]{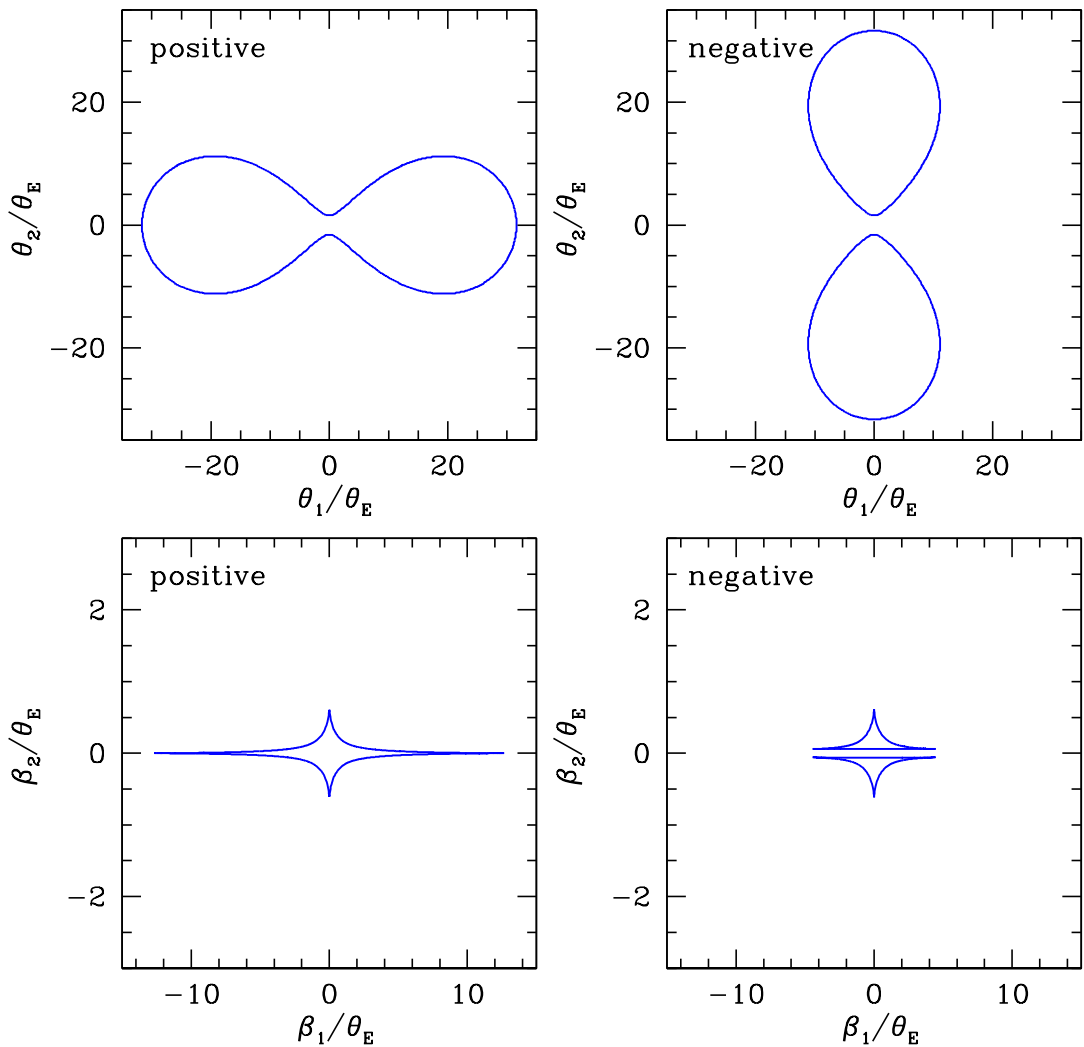}
\end{center}
\caption{Critical curves ({\it upper panels}) and caustics ({\it lower
    panels}) of a point mass lens in the high magnification
  region. {\it Left:} Positive parity case with $\mu_{\rm
    t}^{-1}=0.001$ and $\mu_{\rm r}^{-1}=0.401$.
  {\it Right:} Negative parity case with $\mu_{\rm
    t}^{-1}=-0.001$ and $\mu_{\rm r}^{-1}=0.399$. These are computed
  with {\tt GLAFIC} \cite{Oguri:2010rh}. Note the difference of scales
  between $x$- and $y$-axes in the lower panels.} 
\label{fig:crit}
\end{figure*}

The critical curve can be derived from $\mu^{-1}=0$. Specifically,
\begin{eqnarray}
\left(\frac{\theta}{\theta_{\rm
    E}}\right)^2&=&\frac{\cos(2\phi)}{2}(\mu_{\rm t}-\mu_{\rm
  r})\nonumber\\
  &&\times \left[1\pm\sqrt{1+\frac{4\mu_{\rm t}\mu_{\rm r}}{(\mu_{\rm t}-\mu_{\rm
  r})^2\cos^2(2\phi)}}\right]\nonumber\\
 & \approx & \frac{\mu_{\rm t}\cos(2\phi)}{2}
  \left[1\pm\sqrt{1+\frac{4\mu_{\rm r}({\rm sgn\mu_{\rm
            t}})}{|\mu_{\rm t}|\cos^2(2\phi)}}\right].
  \label{eq:crit}
\end{eqnarray}
A similar equation is also given in \cite{Diego:2017drh}. The caustic
is obtained by converting the critical curve in the image plane to the
source plane via the lens equation. Figure~\ref{fig:crit} shows 
the critical curves and caustics for positive and negative parities,
which have already been given in the literature (e.g., \cite{Chang:1979zz}).

The solutions of equation~(\ref{eq:crit}) depend on the sign of
$\mu_{\rm t}$ (or parity) which is different on each side of the
critical curve.  $\mu_{\rm r}$ maintains the same sign on both sides
of the critical curve and without loss of generality we assume that
$\mu_{\rm r}$ is always positive (i.e, tangential critical curves).  

\subsubsection{Positive parity case: $\mu_{\rm t}>0$}

From equation~(\ref{eq:crit}) we can easily estimate the size of the
critical curve along the $\theta_1$ and $\theta_2$ axes. Along $\theta_1$
(or horizontal axis, i.e., $\cos(2\phi)=1$) we have
\begin{equation}
\theta_{\rm crit,1}\approx \sqrt{\mu_{\rm t}}\theta_{\rm E},
\end{equation}
and along $\theta_2$ (or vertical axis, i.e., $\cos(2\phi)=-1$) we have
\begin{equation}
\theta_{\rm crit,2}\approx \sqrt{\mu_{\rm r}}\theta_{\rm E}.
\end{equation}
Using the lens equation, the corresponding size of the caustic along
$\beta_1$ and $\beta_2$ axes is estimated as
\begin{equation}
\beta_{\rm caus,1}\approx \frac{\sqrt{\mu_{\rm t}}}{\mu_{\rm r}}\theta_{\rm E},
\label{eq:betaplus}
\end{equation}
\begin{equation}
\beta_{\rm caus,2}\approx \frac{1}{\sqrt{\mu_{\rm r}}}\theta_{\rm E}.
\end{equation}

We also want to know the typical width of the caustic. As shown in
Figure~\ref{fig:crit}, it becomes very long and thin. Given that the
size of the critical curves at the region of interest is $\theta_{\rm
  crit}\approx \mathcal{O}(\sqrt{\mu_{\rm t}}\theta_{\rm E})$, from
equation~(\ref{eq:beta2}) we can infer the typical width of the
caustic to (assuming $\sqrt{\mu_{\rm t}}\gg 1$) 
\begin{equation}
\beta_{\rm w}\approx \frac{\theta_{\rm E}}{\sqrt{\mu_{\rm t}}},
\label{eq:deltabeta}
\end{equation}
which implies that the total area enclosed by the caustic is not
significantly enhanced compared with the case of an isolated point
mass lens. Figure~\ref{fig:crit} indicates that the area enclosed by
the caustic at around $\beta_1\approx 0$ has the same order.

\subsubsection{Negative parity case: $\mu_{\rm t}<0$}

In this case, the critical curves do not form along the $\theta_1$ axis
because the right hand side of equation~(\ref{eq:crit}) is always
negative. On the other hand, there are two solutions of
equation~(\ref{eq:crit}) along the $\theta_2$ axis, which we denote
$\theta_{\rm crit,+}$ and  $\theta_{\rm crit,-}$. They are
\begin{equation}
\theta_{\rm crit,+}\approx \sqrt{|\mu_{\rm t}|}\theta_{\rm E},
\end{equation}
\begin{equation}
\theta_{\rm crit,-}\approx \sqrt{\mu_{\rm r}}\theta_{\rm E}.
\end{equation}
The corresponding size of the caustic along the $\beta_2$ axis is
\begin{equation}
\beta_{\rm crit,+}\approx \frac{2}{\sqrt{|\mu_{\rm t}|}}\theta_{\rm E}
\end{equation}
\begin{equation}
\beta_{\rm crit,-}\approx \frac{1}{\sqrt{\mu_{\rm r}}}\theta_{\rm E}.
\end{equation}
We also want to estimate the size of the critical curve and caustic
along the $\theta_1$ direction. First, the maximum $\theta_1$ of the
critical curves is estimated as follows. In the region of interest, we
can approximate $(\theta/\theta_{\rm E})^2\approx \mu_{\rm
  t}\cos(2\phi)$. At the maximum along $\theta_1$,
$\theta_{\rm crit,max}$, $\theta_1^2=\theta^2\cos^2\phi$ must be
stationary, leading to $\cos\phi=1/2$ and $\cos(2\phi)=-1/2$.
Therefore,
\begin{equation}
 \theta_{\rm crit,max}\approx\sqrt{\frac{|\mu_{\rm t}|}{8}}\theta_{\rm E}.
\end{equation}
The corresponding size of the caustic along the $\beta_1$ direction is
\begin{equation}
  \beta_{\rm caus,max}\approx\frac{1}{\mu_{\rm r}}
  \sqrt{\frac{|\mu_{\rm t}|}{8}}\theta_{\rm E}.
\label{eq:betanegative}
\end{equation}
We can use the same argument as above to show that the typical width
of the caustics is given by equation~(\ref{eq:deltabeta}). Thus, the
area enclosed by the caustics is again largely unchanged compared with
the case of an isolated point mass lens, which can also be seen in
Figure~\ref{fig:crit}. By comparing equations~(\ref{eq:betaplus}) and
(\ref{eq:betanegative}), it is found that 
\morv{the length of the caustic is longer in the positive parity than
  in the negative parity by a factor of $2\sqrt{2}$. However,
  Figure~\ref{fig:crit} indicates that in the negative parity there
  are twice more caustic crossings for each lens, which compensates
  the shorter length of the caustic when we estimate the rate of
  caustic crossings (see Section~\ref{sec:eventrate}). }

The source plane region surrounded by $\beta=\pm\beta_{\rm crit,+}$
demagnifies the star. As an example, we consider multiple images for a
source placed at the origin ($\beta_1$, $\beta_2$)=(0, 0). From the lens
equation, we find that there are two multiple images located at
($\theta_1$, $\theta_2$)=($\pm\sqrt{\mu_{\rm r}}\theta_{\rm E}$, 0).
The magnification of the individual images is computed as
\begin{equation}
\mu=-\frac{\mu_{\rm r}^2}{2},
\end{equation}
which is much smaller than the macro magnification $\mu=\mu_{\rm
  t}\mu_{\rm r}$, indicating that any source near the origin are
significantly less magnified compared with the case without the point
mass lens. 

\subsection{Light curves}
\label{sec:lightcurve}

In order to predict light curves we need to assume velocities of
the lens and source with respect to the observer. The transverse
velocity in the lens plane $\mathbf{v}_l$ is converted to the angular
unit as 
\begin{equation}
\mathbf{u}_l = \frac{\mathbf{v}_l}{1+z_l}\frac{1}{D_{ol}},
\end{equation}
where $D_{ol}$ is the angular diameter distance from the observer to
the lens and the factor $1+z_l$ accounts for the dilution of the unit
time, i.e., the angular velocity above indicates the change of the
position on the sky per unit observed-frame time. We divide the lens
velocity into two components, one is a bulk velocity of the whole lens
system (e.g., a peculiar velocity of a galaxy cluster for the case
of MACS~J1149 LS1) and the other is a relative motion of a point mass
lens within the whole lens system. We denote these velocity components
as $\mathbf{v}_m$ and $\mathbf{v}_p$, respectively. We consider these
two components separately as they have different dependences on the
macro lens model  when they are converted to velocities in the source
plane (see below). 

We also consider the transverse velocity in the source plane
$\mathbf{v}_s$. Again, it can be converted to the angular unit as
\begin{equation}
\mathbf{u}_s = \frac{\mathbf{v}_s}{1+z_s}\frac{1}{D_{os}},
\end{equation}
We derive relative velocity of the source and lens in the source plane
by converting the transverse velocity in the lens plane to the source
plane as (see \cite{Kayser:1986aa})
\begin{equation}
  \mathbf{u} =
  \mathbf{u}_m +\left(
    \begin{array}{cc}
     \mu_{\rm r}^{-1} & 0 \\
     0 & \mu_{\rm t}^{-1}
    \end{array}
    \right) \mathbf{u}_p +\mathbf{u}_s.
\label{eq:veltot}    
\end{equation}
This indicates that the relative velocity can be anisotropic. The
magnification tensor comes in because distances between point mass
lenses in the image plane translate into smaller distances in the
source plane due to the cluster potential.
The bulk velocity of a cluster is typically 
$|\mathbf{v}_m|\sim 500~{\rm km\,s^{-1}}$ (e.g.,
\cite{Dolag:2013hj}). Although the relative motion of the point mass
lens, which is of the order of velocity dispersions of massive
clusters, $|\mathbf{v}_p|\sim 1000~{\rm km\,s^{-1}}$, is larger than
the bulk velocity, it is suppressed by the magnification factors as
shown in equation~(\ref{eq:veltot}). 
The contribution from the source motion is expected to be smaller
given the larger redshift and distance to the source. Therefore,
a simple approximation which we adopt in the following discussions is
\begin{equation}
  \mathbf{u} \approx \mathbf{u}_m.
\end{equation}
On the other hand, Figure~\ref{fig:crit} indicates that the caustics
are elongated along the $x$-axis by a factor of $\sqrt{\mu_{\rm t}}\gg
1$. Thus, caustic crossing typically occurs by a source moving along
the $y$-axis with the velocity $\sim |\mathbf{u}_m|$.

When a source crosses the caustics vertically along the $y$-axis, crossing
occurs two times in total for the positive parity case, and four times
for the negative parity case. As discussed above, for the negative
parity case, a source is demagnified near the center.

The behavior of the total magnification near the caustic is important
for estimating the possible maximum magnification. The total
magnification near the caustics is known to behave as $\mu\propto 
\Delta\beta^{-1/2}$, where $\Delta\beta$ is the distance between the
source and the caustic in the source plane. From the analytic
examination and numerical calculations with {\tt GLAFIC}, we find that
the magnifications near the caustics are crudely approximated as
\begin{equation}
\mu(\Delta\beta)\approx \mu_{\rm t}\mu_{\rm r}
\left(\frac{\theta_{\rm E}}{\sqrt{\mu_{\rm
      t}}\Delta\beta}\right)^{1/2},
\label{eq:dbeta}
\end{equation}
which 
\morv{agrees with numerical results within a factor of $\lesssim
  2$. This approximation}
is reasonable given that the width of the caustic is ``shrunk''
by a factor of $\sqrt{\mu_{\rm t}}$ (equation~\ref{eq:deltabeta}).

Equation~(\ref{eq:dbeta}) suggests that the magnification becomes
larger as the source approaches to the caustic, which is a universal
property of lensing near the caustic (see e.g.,
\cite{Blandford:1986zz,Schneider:1992aa}). However, the magnification
saturates when the distance to the caustic becomes comparable to the
size of the source in the source plane, $\beta_R$ (e.g.,
\cite{MiraldaEscude:1991aa}). From this condition, we can estimate the
maximum magnification as  
\begin{equation}
\mu_{\rm max}\approx \mu_{\rm t}\mu_{\rm r}
\left(\frac{\theta_{\rm E}}{\sqrt{\mu_{\rm
      t}}\beta_R}\right)^{1/2}.
\label{sec:mumax}
\end{equation}

\section{Expected properties of the lens and source}
\label{sec:expected}

\subsection{Dependence on the source star}

We discuss how the peak magnification scales with the radius $R$
and luminosity $L$ of a background star. Assuming the black body with
the temperature $T$, they are related as
\begin{equation}
  \frac{L}{L_\odot}=\left(\frac{R}{R_\odot}\right)^2
  \left(\frac{T}{T_\odot}\right)^4.
  \label{eq:blackbody}
\end{equation}
In practice the spectral energy distribution (SED) of a star does not
strictly follow the black body, but this relation still holds
approximately. Using this relation, we find that the observed maximum
flux of the star scales as
\begin{equation}
  f_{\rm max}\propto \mu_{\rm max}L \propto R^{-1/2}L \propto
  R^{3/2}T^4.
  \label{eq:maxflux}
\end{equation}
Therefore, for a given temperature (which can be inferred from the
observation of the SED of the star), the maximum magnified flux of a
larger star is larger than that of a smaller star. 

If the source size is too big compared with the Einstein radius, we do
not observe any microlensing magnifications. From
equation~(\ref{sec:mumax}), we can argue that the source size needs
to satisfy the following condition to have sensitivity to microlensing
\begin{equation}
\sqrt{\mu_{\rm t}}\beta_R \lesssim \theta_{\rm E}.
\label{eq:sizecons1}
\end{equation}
Suppose that we observe a specific microlensing event in which we can
estimate the lower limit of the {\it relative} magnification factor during
the caustic crossing event, $\mu_{\rm obs}$, which is derived from the
difference of the magnitudes measured at the beginning of the event
and when the lensed source is brightest. This gives the lower limit
because the true magnification factor during the caustic crossing
event is larger given the limited sampling and detection limit of
monitoring observations. Thus $\mu_{\rm obs}$ must be smaller than the
relative maximum magnification due to caustic crossing, i.e.,  
$\mu_{\rm obs}<\mu_{\rm max}/(\mu_{\rm t}\mu_{\rm r})$.
Using this condition, we can replace the condition in 
equation~(\ref{eq:sizecons1}) to
\begin{equation}
\sqrt{\mu_{\rm t}}\beta_R \lesssim \frac{\theta_{\rm E}}{\mu_{\rm  obs}^2}.
\label{eq:sizecons2}
\end{equation}
This means that, while the peak brightness is higher for the larger
star simply because of its large intrinsic brightness, the
microlensing magnification is more prominent for the smaller star.

\subsection{Macro model magnification}
\label{sec:macromodel}

The analysis presented in Section~\ref{sec:general} assumed a uniform
macro model magnification for simplicity. In practice, however, the
macro model magnification $\mu_{\rm t}$ and $\mu_{\rm r}$ depends on
the image position with respect to the critical curve of the macro
model. It has been known that $\mu_{\rm t}$ quickly increases as the
image approaches to the critical curve (see, e.g.,
\cite{Blandford:1986zz,Schneider:1992aa}). More specifically, given
that we denote the distance between the image and the critical curve
as $\theta_{\rm h}$, we generally expect $\mu_{\rm t}\propto
\theta_{\rm h}^{-1}$. We parametrize the dependence on $\theta_{\rm h}$
 as 
\begin{equation}
  \mu_{\rm t}(\theta_{\rm h})=\mu_{\rm h}\left(\frac{\theta_{\rm
      h}}{\rm arcsec}\right)^{-1},
\label{eq:mu_th}
\end{equation}
where $\mu_{\rm h}$ is a constant factor that depends on the macro
lens model as well as the location on the critical curve. On other
other hand, $\mu_{\rm r}$ is approximately constant near the critical
curve. 

There is also a well-known asymptotic behavior between $\beta_{\rm h}$
(the angular distance to the caustic in the source plane) and 
$\theta_{\rm h}$ (the angular distance to the critical curve in the
image plane), $\beta_{\rm h}\propto \theta_{\rm h}^2$. In particular,
we parametrize it as
\begin{equation}
  \beta_{\rm h}=\beta_0\left(\frac{\theta_{\rm h}}{\rm
    arcsec}\right)^2.
  \label{eq:mu_beta}
\end{equation}
From these equations we have $\mu_{\rm t}=\mu_{\rm h}(\beta_{\rm
  h}/\beta_0)^{-1/2}$ for the macro model magnification of one of the
merging pair of images.

The maximum magnification (equation~\ref{sec:mumax}) of caustic
crossing is larger for larger $\mu_{\rm t}$, which suggests that stars
that are closer to the critical curve can have higher magnifications. 
However, \cite{Diego:2017drh} (see also \cite{Venumadhav:2017pps})
argued that, even a small fraction of point mass lenses,
significantly changes the asymptotic behavior of the macro model
magnification toward the critical curve. This is because the Einstein
radius of the point mass lens depends on $\mu_{\rm t}$ as $\propto
\sqrt{\mu_{\rm t}}$, and hence for very large $\sqrt{\mu_{\rm t}}$
the Einstein radii for different point mass lenses overlap, even when
the number density of point mass lenses is small. As shown by
ray-tracing simulations in \cite{Diego:2017drh}, beyond this
``saturation'' point the source breaks into many micro-images, and as
a result it loses its sensitivity to the source position with respect
to the macro model caustic. Therefore, the macro model magnification
in fact does not diverge as predicted by equation~(\ref{eq:mu_th}),
but saturates at a finite value. 

To estimate where the saturation happens, \cite{Diego:2017drh}
considered the optical depth $\tau$ defined by
\begin{equation}
\tau=\frac{\Sigma}{M}\pi\left(\sqrt{\mu_{\rm t}}\theta_{\rm E}D_{ol}\right)^2,
\label{eq:tau}
\end{equation}
where $\Sigma$ is the surface mass density of the point mass
component. Here we implicitly assumed that the point masses have the
same mass $M$, although we note that this approximation is reasonably
good when compared with the realistic ray-tracing simulations.
\cite{Diego:2017drh} argued that the saturation happens
when $\tau \approx 1$. From the definition of the Einstein radius
(equation~\ref{eq:einrad}), it is found
\begin{equation}
\tau \propto \mu_{\rm t}\Sigma,
\end{equation}
which indicates that the maximum macro model magnification where the
saturation happens is inversely proportional to the surface mass
density of the point mass component $\Sigma$, and does not depend on
the mass $M$. This means that, in order to achieve high peak
magnifications (equation~\ref{sec:mumax}), lower $\Sigma$ is
preferred. Specifically, we can compute the maximum macro model
magnification by setting $\tau=1$ in equation~(\ref{eq:tau}) as
\begin{equation}
 \mu_{\rm t,max}\approx \frac{M}{\pi\Sigma\left(\theta_{\rm E}D_{ol}\right)^2}.
\label{eq:mutmax1}
\end{equation}
Again, for a fixed surface density $\Sigma$, $\mu_{\rm t,max}$ does
not depend on mass $M$.

We can consider another condition for the saturation from the Einstein
radius (see \cite{Venumadhav:2017pps}). Even for $\tau\ll1$, when the
distance to the macro model critical line, $\theta_{\rm h}$, becomes
comparable to the Einstein radius of the point mass lens, the critical
curves by the point mass lens merge with those from the macro lens
model, and our basic assumption breaks down. Therefore, to have enough
magnifications by the point mass lens, we need the following
condition.
\begin{equation}
\sqrt{\mu_{\rm t}}\theta_{\rm E}\lesssim \theta_{\rm h}.
\end{equation}
Using equation~(\ref{eq:mu_th}), this condition is rewritten as
\begin{equation}
\theta_{\rm E}\lesssim \frac{\mu_{\rm h}}{\mu_{\rm t}^{3/2}}.
\label{eq:temut1} 
\end{equation}
The similar condition in the source plane is
\begin{equation}
\frac{\theta_{\rm E}}{\sqrt{\mu_{\rm t}}}\lesssim \beta_{\rm h},
\end{equation}
which results in
\begin{equation}
\theta_{\rm E}\lesssim \frac{\beta_0\mu_{\rm h}^2}{\mu_{\rm t}^{3/2}}.
\label{eq:temut2} 
\end{equation}
In practice equations~(\ref{eq:temut1}) and (\ref{eq:temut2}) give
quite similar conditions, so in what follows we consider only
equation~(\ref{eq:temut1}). From this condition, we have another
condition for the maximum magnification of the macro mass model
as
\begin{equation}
\mu_{\rm t,max}\approx \left(\frac{\mu_{\rm h}}{\theta_{\rm E}}\right)^{2/3},
\label{eq:mutmax2}
\end{equation}
which indicates $\mu_{\rm t,max} \propto M^{-1/3}$. The true maximum
macro magnification is given by the smaller of $\mu_{\rm t,max}$ given
in equations~(\ref{eq:mutmax1}) and (\ref{eq:mutmax2}).

\subsection{Light curve timescales}

There are important time scales that characterize caustic crossing
events. One is the so-called source crossing time defined by
\begin{equation}
  t_{\rm src}=\frac{2\beta_R}{u},
\label{eq:tsrc}
\end{equation}
where $\beta_R$ is the angular size of the source in the source plane,
and $u$ is the source plane velocity as defined in
equation~(\ref{eq:veltot}). This source crossing time determines the 
timescale of the light curve near the peak. Another important time
scale is given by the time to cross between caustics. Using the width
of the caustics, $\beta_{\rm w}$ (equation~\ref{eq:deltabeta}), it is
expressed as 
\begin{equation}
  t_{\rm Ein}=\frac{\beta_{\rm w}}{u},
\label{eq:tein}
\end{equation}
which gives the typical timescale between multiple caustic crossing
events. The former timescale $t_{\rm src}$ does not depend on the lens
property, whereas the latter timescale $t_{\rm Ein}$ scales with the
lens mass as $\propto M^{1/2}$.

\subsection{Apparent size of microlensed image}

If the Einstein radius is sufficiently large, multiple images can be
resolved. In this case, from the separations of the multiple images we
can directly infer the mass scale of the lens. In usual cases,
however, the Einstein radius of the point mass lens is sufficiently
small compared with the angular resolution of observations. As a
result, the observed microlensed image is point-like, from which we can
set the constraint on the lens mass as
\begin{equation}
\sqrt{\mu_{\rm t}}\theta_{\rm E}\lesssim \sigma_{\theta,{\rm obs}},
\label{eq:appsiz}
\end{equation}
where $\sigma_{\theta,{\rm obs}}$ is the angular resolution of
observations. For the case of {\it Hubble Space Telescope} imaging
observations, we typically have $\sigma_{\theta,{\rm obs}}\approx
0.05$~arcsec. 

\section{Constraining the source and lens properties of MACS~J1149 LS1}
\label{sec:constraint}

We now tune the parameters to the observations of MACS~J1149 LS1
\cite{Kelly:2017fps} and see what kind of constraints we can place on
properties of both the lens object and the source star.

\subsection{Parameters}

The lens redshift of MACS~J1149 LS1 is $z_l=0.544$ and the source
redshift is $z_s=1.49$. \morv{Both the redshifts are spectroscopic
  redshifts and therefore are sufficiently accurate.}
For a cosmology with matter density $\Omega_M=0.3$, cosmological
constant $\Omega_\Lambda=0.7$, and the dimensionless Hubble constant
$h=0.7$, we have 6.4~kpc~arcsec$^{-1}$ at the lens and
8.5~kpc~arcsec$^{-1}$ at the source. The distance modulus to the
source is $45.2$.  With these distances, the Einstein radius of the
(isolated) point mass lens is 
\begin{equation}
  \theta_{\rm E}\approx 1.8\times 10^{-6}
  \left(\frac{M}{M_\odot}\right)^{1/2}{\rm arcsec}.
\label{eq:tein_ls1}
\end{equation}
The angular size of a star as a function of solar radius in the source
plane is
\begin{equation}
\beta_R\approx 2.7\times 10^{-12}\left(\frac{R}{R_\odot}\right){\rm arcsec}.
\label{eq:berls1}
\end{equation}
\morv{We note that uncertainties of our analysis originating from
  cosmological parameter uncertainties are much smaller compared with
  other uncertainties that we will discuss below.}

As discussed in Section~\ref{sec:lightcurve}, we can assume that the
velocity is dominated by that of the bulk motion of the lensing
cluster, $|v|\approx |v_m|\sim 500~{\rm km\,s^{-1}}$. We can convert it to the
angular velocity on the sky as 
\begin{equation}
u \approx 5.2\times 10^{-8} \left(\frac{v}{500~{\rm
  km\,s^{-1}}}\right){\rm arcsec\,yr^{-1}}.
\label{eq:u_ls1}
\end{equation}
\morv{While we fix the $v=500~{\rm km\,s^{-1}}$ in our main analysis,
we also check how the uncertainty on the velocity propagates into
various constraints that we obtain in this paper.}

For the macro mass model using the best-fitting model of the {\tt
  GLAFIC} mass model \cite{Oguri:2010rh,Kawamata:2015haa}, we have
$\mu_{\rm h}\approx 13$ in equation~(\ref{eq:mu_th}) and
$\beta_0\approx 0.045$ in equation~(\ref{eq:mu_beta}). MACS~J1149 LS1
was discovered at $\theta_{\rm h}\approx 0.13$~arcsec at which the
model predict the magnification of $\mu_{\rm t}\approx 100$. On the
other hand, $\mu_{\rm r}\approx 3$ near  MACS~J1149 LS1.
Therefore, the total macro mass model magnification at the position of
MACS~J1149 LS1 is $\mu_{\rm t}\mu_{\rm r}\approx 300$.

\morv{We note that there is uncertainty associated with the macro mass
  model. For example, \cite{Diego:2017drh} noted that {\tt GLAFIC} and
  {\tt WSLAP+} \cite{Diego:2015lla} mass models of MACS~J1149.6+2223
  predicts roughly a factor of 2 different macro model magnifications
  near MACS~J1149 LS1 (see also \cite{Meneghetti:2016hcr} for a test
  of the accuracy of strong lens mass modeling). This difference in
  the macro mass model affects our quantitative results. Again, while
  we fix mass model parameter values to those mentioned above in our
  analysis, we also examine the dependence of our results on macro mass
  model uncertainties by checking the dependence of our results on
  $\mu_{\rm h}$ that differs considerably between  {\tt GLAFIC} and
  {\tt WSLAP+} mass models. } 

The ICL plays a crucial role in the interpretation of the caustic
crossing event. At the position of MACS~J1149 LS1 the surface density
of ICL is estimated as $\Sigma_{\rm ICL}\approx 11-19~M_\odot{\rm
  pc}^{-2}$ depending on assumed stellar initial mass functions
\citep{Kelly:2017fps,Morishita:2017aaa}. Given the critical surface
  density $\Sigma_{\rm crit}\approx 2.4\times 10^3~M_\odot{\rm pc}^{-2}$,
the convergence from the ICL reads $\kappa_{\rm ICL}\approx 0.0046 - 0.0079$. 
This should be compared with the total surface density $\kappa\approx
0.83$ predicted by the best-fitting {\tt GLAFIC} mass model
\cite{Oguri:2010rh,Kawamata:2015haa}. In what follows, we may use the
mass fraction of the point mass lens component defined by
\begin{equation}
f_{\rm p}=\frac{\Sigma}{2000~M_\odot{\rm pc}^{-2}},
\label{eq:fp}
\end{equation}
instead of the surface mass density $\Sigma$.

\subsection{Constraints on MACS~J1149 LS1}
\label{sec:constls1}

Based on discussions given in Section~\ref{sec:expected}, we constrain
the properties of the lens that is responsible for the observed caustic
crossing event, as well as the properties of the source star.

First, from equations~(\ref{sec:mumax}), (\ref{eq:tein_ls1}), and
(\ref{eq:berls1}), at the position of MACS~J1149 LS1 with
 $\mu_{\rm t}\approx 100$ and $\mu_{\rm r}\approx 3$ the maximum
magnification becomes  
\begin{equation}
\mu_{\rm max}\approx 7.8\times 10^4\left(\frac{M}{M_\odot}\right)^{1/4}
\left(\frac{R}{R_\odot}\right)^{-1/2}.
\label{eq:mumax_ls1}
\end{equation}
Assuming the temperature to be $T=12000~K$ (see \cite{Kelly:2017fps}),
the absolute $V$-band magnitude of the star before the magnification
is computed using equation~(\ref{eq:blackbody}) as
\begin{equation}
M_{\rm star}\approx 2.2-5\log\left(\frac{R}{R_\odot}\right),
\end{equation}
where we used $M_{L_\odot}\approx 4.8$ in $V$-band, $T_\odot=5777~K$,
and included the bolometric correction. Therefore, taking account of
the cross-filter K-correction derived in \cite{Kelly:2017fps}, the
minimum apparent magnitude of the star magnified by microlensing at
the peak is 
\begin{eqnarray}
m_{\rm peak}&\approx&46.4-5\log\left(\frac{R}{R_\odot}\right)-2.5\log\mu_{\rm
  max},\nonumber\\
&\approx& 34.1-3.75\log\left(\frac{R}{R_\odot}\right)-0.625
\log\left(\frac{M}{M_\odot}\right).
\label{eq:mmin}
\end{eqnarray}
The observation indicates that the peak magnitude is brighter than
$m=26$ \cite{Kelly:2017fps}, i.e.,  
\begin{equation}
3.75\log\left(\frac{R}{R_\odot}\right)+0.625
\log\left(\frac{M}{M_\odot}\right)\gtrsim 8.1.
\label{eq:const1}
\end{equation}
During the caustic crossing event the source star is magnified at
least a factor of 3 or so. Setting $\mu_{\rm obs}\approx 3$ in
equation~(\ref{eq:sizecons2}), we obtain constraint on the source size
as 
\begin{equation}
\left(\frac{R}{R_\odot}\right)\left(\frac{M}{M_\odot}\right)^{-1/2}
\lesssim 7600.
\label{eq:const2}
\end{equation}

We now consider the conditions that the saturation does not
happen, because the quantitative constraints derived above assumed
no saturation at the position of MACS~J1149 LS1. By setting $\mu_{\rm
  t}=100$ in equation~(\ref{eq:mutmax1}), we obtain
\begin{equation}
\Sigma\simlt 24~M_\odot{\rm pc}^{-2}.
\label{eq:const3}
\end{equation}
Again, we caution that this is the result assuming that point mass
lenses have the same mass $M$. As shown above, the ICL component has
the surface density of 
$\Sigma_{\rm ICL}\approx 11-19~M_\odot{\rm pc}^{-2}$ (which corresponds to
the mass fraction of $f_{\rm ICL}=\Sigma_{\rm ICL}/\Sigma_{\rm
  tot}\approx 0.0055-0.0095$), and therefore satisfy this
condition. On the other hand, from the other condition for
non-saturation (equation~\ref{eq:mutmax2}) we have
\begin{equation}
M\lesssim 5.1\times 10^7M_\odot.
\label{eq:const4}
\end{equation}

For MACS~J1149 LS1, the source crossing time (equation~\ref{eq:tsrc}) is
\begin{equation}
  t_{\rm src}\approx 0.038
  \left(\frac{R}{R_\odot}\right) \left(\frac{v}{500~{\rm
  km\,s^{-1}}}\right)^{-1}{\rm days}.
\end{equation}
Similarly, the time scale between caustic crossings is
\begin{equation}
  t_{\rm Ein}\approx
3.5\left(\frac{M}{M_\odot}\right)^{1/2}
\left(\frac{v}{500~{\rm km\,s^{-1}}}\right)^{-1} {\rm yr}.
\label{eq:tein2}
\end{equation}
In the case of MACS~J1149 LS1, the source crossing time, which is the
timescale where the light curve is affected by the finiteness of the
source star radius very near the peak, appears to be smaller than
$\sim 10$~days \cite{Kelly:2017fps}. The condition 
$t_{\rm src}\lesssim 10$~days becomes  
\begin{equation}
\left(\frac{R}{R_\odot}\right) \left(\frac{v}{500~{\rm
  km\,s^{-1}}}\right)^{-1}\lesssim 260.
\label{eq:const5}
\end{equation}
For this source size, the apparent magnitude of the star without
microlensing by the point mass lens is $\lesssim 28$, which appears to
be consistent with the observation \cite{Kelly:2017fps}.
Also $t_{\rm Ein}$ seems to be at least larger than $\sim 1$~yr, so
$t_{\rm Ein}\gtrsim 1$~yr gives rise to
\begin{equation}
  \left(\frac{M}{M_\odot}\right)\left(\frac{v}{500~{\rm
      km\,s^{-1}}}\right)^{-2}\gtrsim 0.082.
\label{eq:const6}
\end{equation}

Since MACS~J1149 LS1 was unresolved during the caustic crossing event,
we use equation~(\ref{eq:appsiz}) to set constraint on the mass of the
lens as
\begin{equation}
M\lesssim 7.6\times 10^6M_\odot.
\label{eq:const7}
\end{equation}
By using this argument we may also exclude the possibility that the
caustic crossing event was produced by massive dark matter
substructures.

\begin{figure}[t]
\begin{center}
\includegraphics[width=0.45\textwidth]{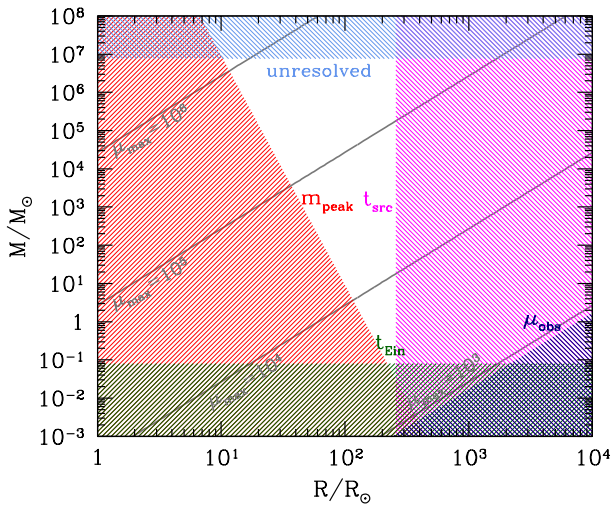}
\end{center}
\caption{Constraints on the source radius ($R$) and lens mass
  ($M$) for MACS~J1149 LS1. Shaded regions show excluded regions from
  various constrains. Specifically, we consider constraints from the
  peak magnitude ($m_{\rm peak}$, equation~\ref{eq:const1}), the
  magnification during the caustic crossing ($\mu_{\rm obs}$,
  equation~\ref{eq:const2}), the source crossing time ($t_{\rm src}$,
  equation~\ref{eq:const5}), the time scale between caustic crossings
  ($t_{\rm Ein}$, equation~\ref{eq:const6}), and the unresolved shape
  during caustic crossing (unresolved, equation~\ref{eq:const7}). The
  saturation condition given by equation~(\ref{eq:const4}) is always
  satisfied in the allowed region of this plot. Contours show the
  constant peak magnification (equation~\ref{eq:mumax_ls1}) in this
  parameter space. From top to bottom, we show contours of $\mu_{\rm
    max}=10^6$, $10^5$, $10^4$, and $10^3$.}
\label{fig:const}
\end{figure}

We now put together all these constraints and derive allowed ranges of
the lens mass $M$ and the source size $R$. The result is shown in
Figure~\ref{fig:const}, where we fixed the bulk velocity of the lens
to $v=500~{\rm km\,s^{-1}}$. We find that there are large ranges of
the lens mass and the source size that can explain MACS~J1149 LS1. 

However, it is expected that the lens and source populations are not
distributed uniformly in this parameter space. As discussed in the
next Section, the size distribution of the source star is expected to
be significantly bottom-heavy, i.e., stars with smaller radii are more
abundant than those with larger radii. The same argument also holds
for the lens mass, if we assume standard stars and stellar remnants as
the lens population, but with the minimum mass of $\approx 0.3~M_\odot$
below which the stellar initial mass function is truncated. 
Therefore, in the allowed parameter space, the
most likely set of parameters are $R\approx 180~R_\odot$ and $M\approx 
0.3~M_\odot$. In this case, the star is magnified by a factor of
$\approx 4300$ at the peak. The result is fully consistent with the
scenario that a blue supergiant is magnified by a foreground ICL star.

On the other hand, our result does not exclude the possibility that
the microlensing is caused by an exotic population such as PBHs with
masses between $\sim 1~M_\odot$ and $\sim 10^6~M_\odot$, as long as
they have low surface density so that they satisfy the saturation
condition. For more massive lenses, Figure~\ref{fig:const} indicates
that the peak magnification is even higher and can reach up to 
$\sim 10^4-10^6$.  

\section{Event rates}
\label{sec:eventrate}

\subsection{Star population in the arc}

It is estimated that the surface brightness of the arc is $\approx
25$~mag~arcsec$^{-2}$ in F125W band \cite{Kelly:2017fps}, which
corresponds to $\approx 6.5 \times 10^9L_\odot$~arcsec$^{-2}$. We need to
convert the observed arc surface brightness to the number density of
stars that can be magnified by caustic crossing events. We do so by
simply assuming a power-law luminosity function of stars,
$dn/dL\propto L^{-2}$, as considered in \cite{Kelly:2017fps}. 
The normalization of the luminosity function is determined so that
that the total luminosity density $\int^{L_{\rm max}}_{L_{\rm
    min}}L(dn/dL)dL$ matches the observed surface brightness. 
Assuming the luminosity range of $L_{\rm min}=0.1L_\odot$ and $L_{\rm
  max}=10^7L_\odot$, the surface number density can be converted to
the number density of stars in the image plane
\begin{eqnarray}
n_{\rm star}(L_1<L<L_2)&= &\frac{6.5 \times 10^9\,{\rm arcsec^{-2}}}{\mu_{\rm t}\mu_{\rm r}\ln(L_{\rm
  max}/L_{\rm
    min})}\left(\frac{L_\odot}{L_1}-\frac{L_\odot}{L_2}\right)\nonumber\\
& & \hspace*{-1.0cm}
\approx \frac{3.5\times 10^8\,{\rm arcsec^{-2}}}{\mu_{\rm t}\mu_{\rm r}}
\left(\frac{L_\odot}{L_1}-\frac{L_\odot}{L_2}\right),
\end{eqnarray}
where $\mu_{\rm t}\mu_{\rm r}$ in the denominator accounts for the
lensing magnification of luminosities of individual stars. Using
equation~(\ref{eq:blackbody}) and fixing $T=12000~K$, we can convert
this to the number density in the star radius range
\begin{equation}
n_{\rm star}(R_1<R<R_2)=\frac{n_0}{\mu_{\rm t}\mu_{\rm r}}
\left[\left(\frac{R_\odot}{R_1}\right)^2-\left(\frac{R_\odot}{R_2}\right)^2\right],
\end{equation}
where $n_0=1.9\times 10^7\,{\rm arcsec^{-2}}$.
For a given lens mass $M$, the lower limit of the radius comes from
the constraint on the peak magnitude. Here we consider caustic crossing
events with $m_{\rm peak}<26$. From equation~(\ref{eq:mmin})
\begin{equation}
R_1\approx 140\left(\frac{\mu_{\rm t}}{100}\right)^{-0.5}
\left(\frac{M}{M_\odot}\right)^{-0.167}R_\odot,
\end{equation}
where we recover the dependence on $\mu_{\rm t}$ which originates from
equation~(\ref{sec:mumax}). We set $R_2\approx 730\,R_\odot$, 
the radius corresponding to $L_{\rm max}=10^7L_\odot$.

\begin{figure}[t]
\begin{center}
\includegraphics[width=0.45\textwidth]{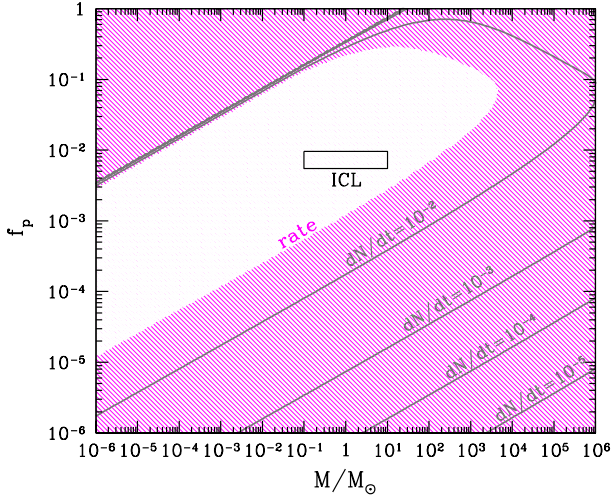}
\end{center}
\caption{Constraints in the $M$-$f_{\rm p}$ plane for MACS~J1149 LS1,
  where $M$ is the lens mass and $f_{\rm p}$ is the mass fraction of
  the point mass component to the total mass. Shaded regions show
  excluded regions from the event rate (rate,
  equation~\ref{eq:constn}). The small rectangular region 
  shows the rough mass fraction and the mass range of ICL stars.
  Contours show the constant even rate in this plane. From inner to
  outer contours, we show contours for $dN/dt=10^{-2}$,
  $10^{-3}$, $10^{-4}$, and $10^{-5}$.}
\label{fig:rate}
\end{figure}

\subsection{Expected rate}
\label{sec:erate}

From the analysis in Section~\ref{sec:general}, we know that the
typical length scale of the caustic along the $\beta_1$ direction is
$\approx 2(\sqrt{\mu_{\rm t}}/\mu_{\rm r})\theta_{\rm E}$. Therefore
the expected event rate is 
\begin{eqnarray}
\frac{dN}{dt}&=&2\int_{\theta_{\rm h,min}}^{\theta_{\rm h,max}}
d\theta_{\rm h} n_{\rm star}w_{\rm arc}\mu_{\rm t}\mu_{\rm r}
\frac{\Sigma}{M}2\frac{\sqrt{\mu_{\rm t}}}{\mu_{\rm
    r}}D_{ol}^2\theta_{\rm E}u,\nonumber\\
&&+2\int_0^{\theta_{\rm h,min}}d\theta_{\rm h}n_{\rm star}w_{\rm arc}\mu_{\rm t}\mu_{\rm r}
\frac{\Sigma}{M}2\left.\frac{\sqrt{\mu_{\rm t}}}{\mu_{\rm
    r}}D_{ol}^2\theta_{\rm E}u\right|_{\mu_{\rm t,max}}\nonumber\\
&=&2\int_{\mu_{\rm t,min}}^{\mu_{\rm t,max}} \frac{\mu_{\rm h}d\mu_{\rm t}}{\mu_{\rm
    t}^2}n_0\left[\left(\frac{R_\odot}{R_1}\right)^2-\left(\frac{R_\odot}{R_2}\right)^2\right]
\nonumber\\
&&\times w_{\rm arc}\frac{\Sigma}{M}2\frac{\sqrt{\mu_{\rm t}}}{\mu_{\rm r}}
D_{ol}^2 \theta_{\rm E}\nonumber\\
&&+2\theta_{\rm h,min}n_{\rm star}w_{\rm arc}\mu_{\rm t}\mu_{\rm r}
\frac{\Sigma}{M}2\left.\frac{\sqrt{\mu_{\rm t}}}{\mu_{\rm
    r}}D_{ol}^2\theta_{\rm E}u\right|_{\mu_{\rm t,max}},
\label{eq:dndt_full}
\end{eqnarray}
where $w_{\rm arc}$ (assumed to be $0.2$~arcsec in the following
calculations) is the width of the giant arc along the critical curve,
and the saturation conditions give the upper limit $\mu_{\rm t,
  max}$. The factor $\mu_{\rm t}\mu_{\rm r}$ converts the number
density of the point mass lens in the image plane, $\Sigma/M$, to the
corresponding number density in the source plane. The prefactor $2$ is
introduced due to the fact that caustic crossing events can happen on
both sides of the critical curve. As shown in
Section~\ref{sec:general}, while the length of the caustic is shorter
in the negative parity region, there are twice more caustic crossings
for each lens, which would compensate the shorter length of the caustic.
The second term represents the contribution from the saturate region
in which caustic crossing events 
are observed (see \cite{Diego:2017drh,Venumadhav:2017pps}). We make a
simple assumption that the rate calculation of the saturated region is
same as that for the unsaturated region but with replacing $\mu_{\rm
  t}$ to the saturation value $\mu_{\rm t, max}$. Among the saturation
conditions given in  equations~(\ref{eq:mutmax1}) and
(\ref{eq:mutmax2}), parameter values of $\Sigma$ (or equivalently
$f_{\rm p}$ defined in equation~\ref{eq:fp}) and $M$ determines which
condition determine the maximum $\mu_{\rm t}$. These two conditions
reduce to 
\begin{equation}
\mu_{\rm t, max}=1.2f_{\rm p}^{-1},
\label{eq:mitmax3}
\end{equation}
\begin{equation}
\mu_{\rm t, max}=3.7\times 10^4 \left(\frac{M}{M_\odot}\right)^{-1/3}.
\label{eq:mitmax4}
\end{equation}
By equating these two conditions we can define the critical point mass
fraction $f_{\rm p, crit}$
\begin{equation}
f_{\rm p, crit}=3.2\times 10^{-5}\left(\frac{M}{M_\odot}\right)^{1/3}.
\end{equation}
When $f_{\rm p}>f_{\rm p, crit}$ $\mu_{\rm t, max}$ is determined from
equation~(\ref{eq:mitmax3}), whereas when $f_{\rm p}<f_{\rm p, crit}$
$\mu_{\rm t, max}$ is determined from equation~(\ref{eq:mitmax4}).
One condition to determine $\mu_{\rm t,min}$ is $R_1=R_2$, which gives
$\mu_{\rm t,min}\lesssim 1$. In practice, $\mu_{\rm t,min}$ would be
determined by the extent of the arc in the direction perpendicular to
the critical curve. We tentatively set $\mu_{\rm t,min}=10$, which
correspond to the maximum distance from the critical curve,
$\theta_{\rm h,max}\approx 1.3''$ (see equation~\ref{eq:mu_th}). In
some cases, however, $\mu_{\rm t,min}$ determined from $R_1=R_2$
because larger than $10$, and in that case we adopt the former value
as $\mu_{\rm t,min}$. 

\begin{figure}[t]
\begin{center}
\includegraphics[width=0.45\textwidth]{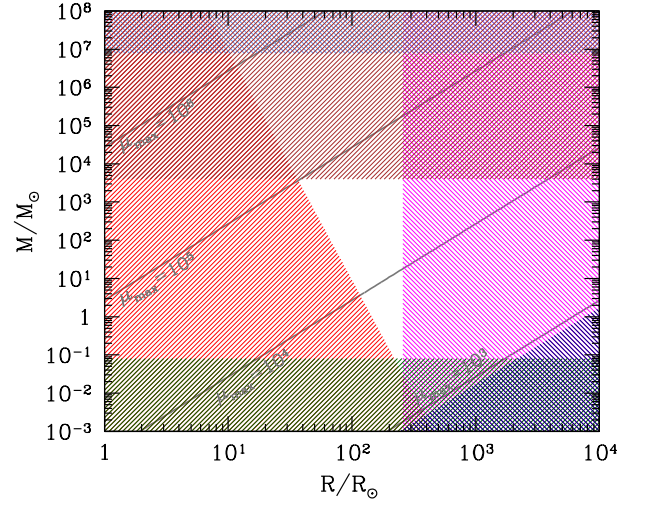}
\end{center}
\caption{Similar to Figure~\ref{fig:const}, but additional constraint
  on the lens mass range from the event rate (see
  Figure~\ref{fig:rate}) is included. }
\label{fig:const_rate}
\end{figure}

Plugging in the parameter values for MACS~J1149 LS1, we have
\begin{eqnarray}
\frac{dN}{dt}&\approx &5.0\times 10^5\,f_{\rm p}\left(\frac{M}{M_\odot}\right)^{-1/2}
\left(\frac{v}{500~{\rm km\,s^{-1}}}\right)I(M)\nonumber\\
&&+4.5\times 10^4\,f_{\rm p}\left(\frac{M}{M_\odot}\right)^{-1/2}
\left(\frac{v}{500~{\rm km\,s^{-1}}}\right)J(M),
\label{eq:dndt}
\end{eqnarray}
\begin{eqnarray}
I(M)&=&\int_{\mu_{\rm t,min}}^{\mu_{\rm t,max}} \frac{d\mu_{\rm t}}{\mu_{\rm
    t}^{3/2}}\left[\left(\frac{R_\odot}{R_1}\right)^2-\left(\frac{R_\odot}{R_2}\right)^2\right]
\nonumber\\
&\approx& 9.3\times 10^{-7}\left(\mu_{\rm t,max}^{1/2}-\mu_{\rm
  t,min}^{1/2}\right)\left(\frac{M}{M_\odot}\right)^{1/3}\nonumber\\
&& +3.7\times 10^{-6}\left(\mu_{\rm t,max}^{-1/2}-\mu_{\rm t,min}^{-1/2}\right),
\end{eqnarray}
\begin{eqnarray}
J(M)&=& \left(\frac{R_\odot}{\left.R_1\right|_{\mu_{\rm t,max}}}\right)^2-\left(\frac{R_\odot}{R_2}\right)^2
\nonumber\\
&\approx& 4.7\times 10^{-7}\mu_{\rm t, max}\left(\frac{M}{M_\odot}\right)^{1/3}-1.9\times 10^{-6}
\end{eqnarray}
where $dN/dt$ is the event rate, i.e., the number of caustic crossing
events per year.

We compute the event rate as a function of lens mass $M$ and mass
fraction of the lens $f_{\rm p}$ using equation~(\ref{eq:dndt}). We
can use this predicted rate calculation to place additional constraints
on the lens population. Since MACS~J1149 LS1 is observed with $\sim 2$
year monitoring observations of MACS~J1149, the $2\sigma$ limit of
the predicted rate is  
\begin{equation}
\frac{dN}{dt}\gtrsim 0.025~{\rm year}^{-1},
\label{eq:constn}
\end{equation}
here we do not consider the additional event (Iapyx) in the rate
constraint because its peak brightness may be fainter than $26$~mag. 

Figure~\ref{fig:rate} shows the constraint from the event rate
(equation~\ref{eq:constn}) in the $M$-$f_{\rm p}$ plane. 
We find that there exists an allowed region with $M\lesssim 
4\times 10^3M_\odot$ and $f_{\rm p}\lesssim 0.3$. 
For a fixed $f_{\rm p}$, large lens masses result in low event rates
because the mean free path of a source is proportional to $M^{-1/2}$,
whereas small lens masses also result in low event rates because of
the lower maximum magnification (see equation~\ref{eq:mumax_ls1}).
Interestingly, this allowed region is fully consistent with the ICL
component which has the right mass range and mass fraction. Therefore,
together with the result shown in Figure~\ref{fig:const}, we conclude
that the observation of MACS~J1149 LS1 is fully explained by
microlensing due to an ICL star.

When marginalized over $f_{\rm p}$, the result in
Figure~\ref{fig:rate} provides additional constraint on the mass range
of the point lens component that produced MACS~J1149 LS1. Thus, in
Figure~\ref{fig:const_rate}, we revisit the constraints in the $R$-$M$
plane, with the additional constraint from Figure~\ref{fig:rate}. We
find that the additional constraint makes the allowed ranges of the
lens mass and source size narrower. However, the most plausible values
of $R\approx 180~R_\odot$ and $M\approx 0.3~M_\odot$ assuming the
prior distributions (see discussions in Section~\ref{sec:constls1}) are
kept unchanged by this additional constraint. This new constraint from
the event rate limits more severely the possibility of explaining
MACS~J1149 LS1 by exotic dark matter models such as PBHs.

\begin{figure}[t]
\begin{center}
\includegraphics[width=0.45\textwidth]{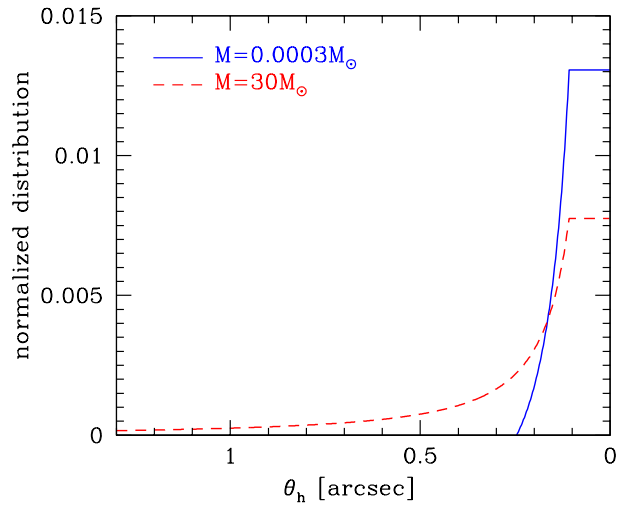}
\end{center}
\caption{The expected distribution of the positions of caustic
  crossing events from the macro model critical curve $\theta_{\rm
    h}$, i.e., the normalized differential distribution of the event
  rate, $(d^2N/d\theta_{\rm h}dt)/(\int d\theta_{\rm h}
  d^2N/d\theta_{\rm h}dt)$.  The distribution is essentially the
  integrand of equation~(\ref{eq:dndt_full}). Parameters are tuned for
  those of MACS~J1149 LS1 as considered in Section~\ref{sec:constls1}. We
  consider two different masses of the point mass lens component,
   $M=0.0003~M_\odot$ ({\it solid}) and $30~M_\odot$ ({\it
    dashed}). Given the relation given in equation~(\ref{eq:mu_th}),
  the distribution of $\theta_{\rm h}$ can also be converted to that of
   the macro model magnification $\mu_{\rm t}$.}
\label{fig:mudist}
\end{figure}

Are there any ways to further constrain the lens mass? One possible
way is to check the positions of caustic crossing events. Because
point mass lenses with larger masses can produce higher magnifications
due to caustic crossing, the macro lens magnification $\mu_{\rm t}$
required to exceed the peak magnitude threshold can be smaller. This
means that, microlensing by large mass lenses can be observed at
positions further away from the critical curve of the macro mass
model. We check this point explicitly by computing distributions of
the distance from the macro model critical curve, $\theta_{\rm h}$,
for different lens masses. The result shown in
Figure~\ref{fig:mudist}, which plots the normalized differential
distribution of the event rate as a function of $\theta_{\rm h}$,
indicates that this is indeed the case. With just one event it is
impossible to conclude which lens mass is favored, but by observing
many caustic crossing events near the critical curve we may be able to
constrain the mass of the point mass lens component more
directly.

Another way to better constrain the lens mass is to observe multiple
caustic crossing events by a single lens, because the time interval
between those events provide information on the mass of the lens (see
equation~\ref{eq:tein}). For instance, the light curve of MACS~J1149
LS1 shows a possible peak in the spring of 2014
\citep{Kelly:2017fps}. If this is interpreted as a caustic crossing
event from the same lens, the observation suggests $t_{\rm Ein}\approx
2$~years, which implies $M\approx 0.3~M_\odot$ from
equation~(\ref{eq:tein2}), which is fully consistent with other
constraints derived in this paper. 

\begin{figure}[t]
\begin{center}
\includegraphics[width=0.45\textwidth]{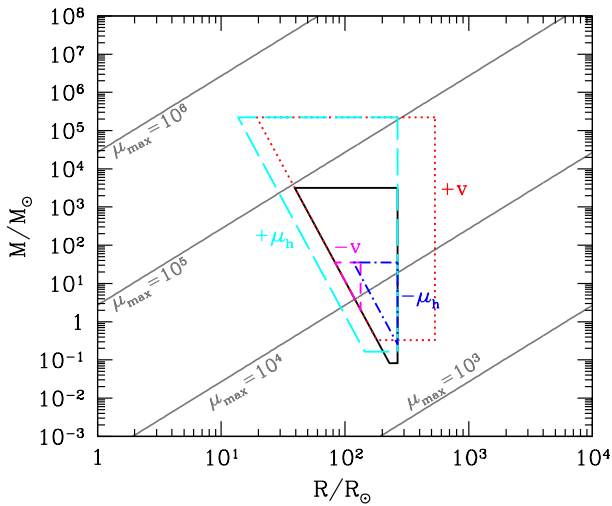}
\end{center}
\caption{\morv{Dependence of constraints in the $R$-$M$ plane shown in
  Fig.~\ref{fig:const_rate} on model parameters. The solid line
  indicates an allowed region in the $R$-$M$ plane in our fiducial
  setup, as shown in Fig.~\ref{fig:const_rate}. Dotted and short
  dashed lines show how the allowed region changes by changing the
  transverse velocity $v$ by factors of $2$ and $0.5$,
  respectively. Long dashed and dot-dashed lines show  how the allowed
  region changes by changing the macro model magnification $\mu_{\rm
    h}$ by factors of $2$ and $0.5$, respectively.}}
\label{fig:consterr}
\end{figure}

\subsection{Effects of model parameter uncertainties}
\label{sec:uncertainty}

\morv{Predictions of event rates and constraints on lens and source
  properties described in the previous subsection are subject to model
  parameter uncertainties. Among others, the assumption on the
  transverse velocity is a source of large uncertainties given that
  the probability distribution of the velocity is quite broad with the
  width of the distribution being a factor of $\sim 2$
  \cite{Dolag:2013hj}. In addition, as discussed in
  Section~\ref{sec:constraint}, the macro model magnification may also
  involve large uncertainty, which can have large impact on our
  results. Here we explore effects of model parameter uncertainties on
  our results, focusing on uncertainties associated with the
  transverse velocity $v$ and the macro model magnification $\mu_{\rm
    h}$ defined in equation~(\ref{eq:mu_th}), assuming the uncertainty
  of $\pm 0.5$~dex (i.e., factors of $2$ and $0.5$) on these
  parameters. }

\morv{First, we discuss uncertainties of predicted event rates due to
  uncertainties of $v$ and $\mu_{\rm h}$. From
  equation~(\ref{eq:dndt_full}), it is found that the event rate is
  linearly proportional to both $v$ and $\mu_{\rm h}$. Therefore, the
  uncertainty of $\pm 0.5$~dex on $v$ and $\mu_{\rm h}$ directly
  translates into the uncertainty of $\pm 0.5$~dex on the predicted
  event rate. This indicates that our prediction on the event rate for
   Icarus is uncertainty by a factor of 2 or so. }

\morv{Next, we check the effects of the uncertainty on these
  parameters on our constraints on lens and source properties that
  are summarized in Fig.~\ref{fig:const_rate}. In
  Fig.~\ref{fig:consterr}, we show how the allowed region shown in
  Fig.~\ref{fig:const_rate} changes by shifting $v$ and $\mu_{\rm h}$
  by $\pm 0.5$~dex. We find that the impact of these model parameter
  uncertainty on our results is indeed significant. Interestingly, the
  lens mass of $\sim 1~M_\odot$ and the source star radius of $\sim
  100~R_\odot$ is allowed even if we take account of these model
  parameter uncertainties.
}

\begin{figure}[t]
\begin{center}
\includegraphics[width=0.37\textwidth]{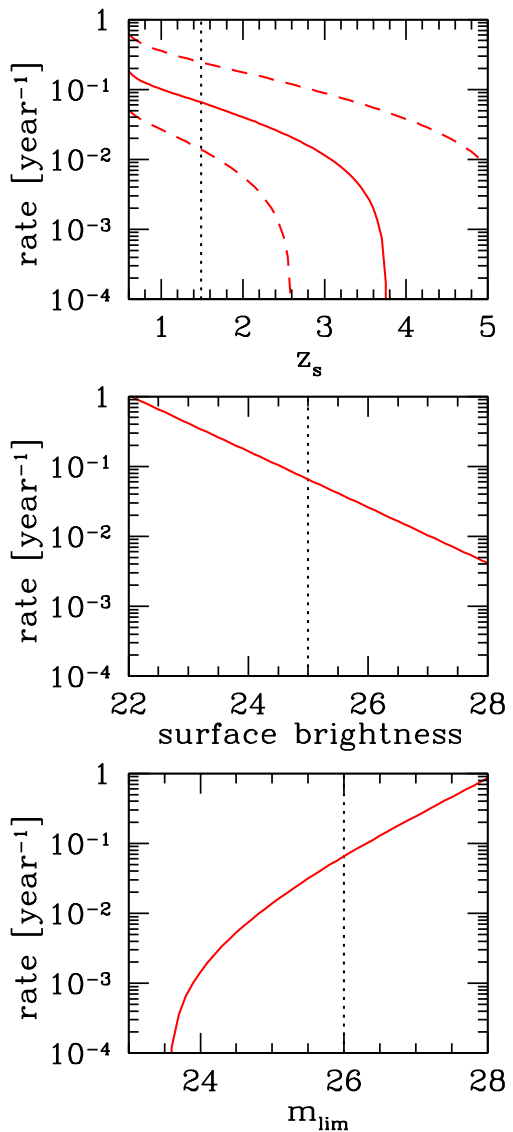}
\end{center}
\caption{The dependences of the event rate
  (equation~\ref{eq:dndt_full}) on various model parameters. We
  consider model parameters that are tuned for the MACS~J1149 LS1 (see
  Section~\ref{sec:constls1}, and fix the mass fraction of the point
  mass lens $f_{\rm p}=0.01$ and the lens mass $M=0.3~M_\odot$. From
  top to bottom panels, we show the dependences of the event rate on
  the source redshift $z_s$, the surface brightness of the arc, and
  the limiting magnitude of the monitoring observation $m_{\rm lim}$.
  The vertical dotted lines show our fiducial values for MACS~J1149
  LS1. In the top panel, we also show the results for the limiting
  magnitudes brighter and fainter by 1~mag (i.e., $m_{\rm lim}=25$
  and $27$) by dashed lines. }
\label{fig:rate_par}
\end{figure}

\subsection{Dependence of event rates on model parameters}

The expected event rate given by equation~(\ref{eq:dndt_full}) depends
on various model parameters. To understand the model dependence on the
event rate, we repeat the computation of the event rate for MACS~J1149
LS1, changing one of the model parameters while fixing the other model
parameters. Here we change the source redshift $z_s$, the surface
brightness of the arc, and the limiting magnitude of the monitoring
observation $m_{\rm lim}$. We show the result in
Figure~\ref{fig:rate_par}. 

We find that the event rate is relatively a steep function of the source
redshift $z_s$. This is simply because we need higher magnification
in order for stars at higher redshifts to be observed. Since the
observed maximum flux is an increasing function of the source radius
$R$ (see equation~\ref{eq:maxflux}), stars detected in giant arcs at
higher redshifts correspond to intrinsically more luminous stars. The
event rate becomes zero beyond $z_s\sim 3.8$, because there is no star
that has the observed maximum flux that exceeds the detection limit
$m_{\rm lim}$.

The dependence on the surface brightness is easily understood. The
number of stars is proportional to the total luminosity of the arc,
and increasing the surface brightness with fixed arc size simply
increases the total luminosity.

We also find that the dependence of the event rate on the limiting
magnitude $m_{\rm lim}$ is strong. For example, monitoring with 2
magnitude deeper images, which can be enabled with James Webb Space
Telescope,  may be able to detect $\sim 10$ times more caustic
crossing events, allowing more detailed statistical studies of the
caustic crossing events such as the spatial distribution as discussed
in Section~\ref{sec:erate}. Again, equation~(\ref{eq:maxflux})
indicates that deeper observations allow us to detect less luminous
stars, which are more abundant. There is a sharp cutoff at
$m_{\rm  lim}\sim 23.5$ by the same reason as in the source redshift $z_s$.

\section{Constraints on compact dark matter in the presence of ICL}
\label{sec:compactdm}

As discussed in \cite{Kelly:2017fps,Diego:2017drh,Venumadhav:2017pps},
even though ICL stars can fully explain MACS~J1149 LS1, we can still
place constraints on compact dark matter scenario where some fraction
of dark matter consists of compact objects such as black holes. 
This is because such compact dark matter can break the caustic due to
the macro lens model into micro-caustics, which reduce the
magnification significantly (see discussions in
Section~\ref{sec:macromodel}). The high fraction of compact dark
matter leads to significant saturation at the position of MACS~J1149
LS1, which effectively reduces the macro model magnification at that
position. Since the smaller macro model magnification leads to fainter
peak magnitudes of caustic crossing events, the high level of
saturation can be inconsistent with the observation of MACS~J1149 LS1.

We can quantify the constraint as follows. From the peak magnitude
(equation~\ref{eq:mmin}) and the constraint on the source radius
(equation~\ref{eq:const5}), we can derive minimum (brightest) peak
magnitude as 
\begin{eqnarray}
m_{\rm peak,min}&\approx&
25.1-0.625\log\left(\frac{M}{M_\odot}\right)\nonumber\\
&&-1.875\log\left(\frac{\mu_{\rm t, LS1}}{100}\right),
\end{eqnarray}
where $\mu_{\rm t, LS1}$ is given by equation~(\ref{eq:mitmax3}) for
the case of interest here. For compact dark matter with masses
$M<10~M_\odot$, we conservatively assume that the MACS~J1149 LS1 is
produced by an ICL star with the mass $10~M_\odot$ because larger
lens masses predict brighter peak magnitudes. Given the
condition $m_{\rm  peak,min}<26$ we obtain the following constraint on
$f_{\rm p}$ 
\begin{eqnarray}
f_{\rm p}<0.08.
\label{eq:pbhconst1}
\end{eqnarray}
For compact dark matter with masses $M>10~M_\odot$ we can achieve
brighter peak magnitudes by assuming that the caustic crossing event
was produced by compact dark matter rather than an ICL star. In this
case, from the same condition $m_{\rm peak,min}<26$ we obtain
\begin{eqnarray}
\log f_{\rm p}<-1.44+\frac{1}{3}\log\left(\frac{M}{M_\odot}\right).
\end{eqnarray}
The condition given in equation~(\ref{eq:pbhconst1}) is in principle
independent of the mass of the compact dark matter microlens, which
means that this constraint can be applied for a wide mass range below
$10~M_\odot$. However, when the mass of the compact dark matter
microlens is very small, the extent of the caustic produced by compact
dark matter becomes much smaller than the source size. In this case,
any lensing effects by compact dark matter is smeared out due to the
finite source size effect, and as a result it does not cause any
saturation. We can write this condition as 
\begin{equation}
\frac{\theta_{\rm E}}{\sqrt{\mu_{\rm t}}}\lesssim \beta_R.
\end{equation}
Given the allowed range of the source radius $R$ and $\mu_{\rm
  t}<100$, this condition reduces to 
\begin{equation}
M\lesssim 1.5\times 10^{-5}M_\odot.
\end{equation}

\begin{figure}[t]
\begin{center}
\includegraphics[width=0.45\textwidth]{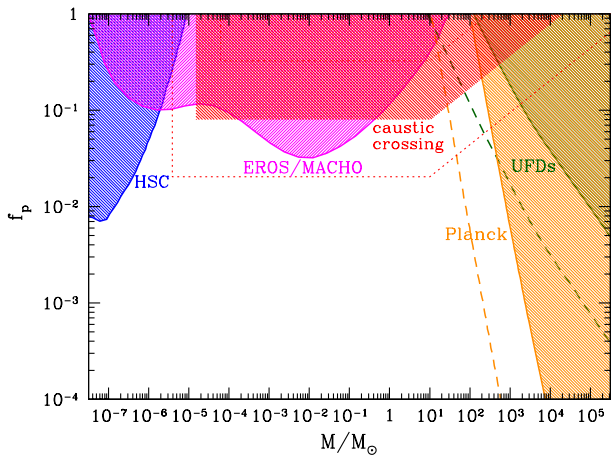}
\end{center}
\caption{Constraints on the mass ($M$) and abundance ($f_{\rm p}$) of
  compact dark matter. Shaded regions show excluded regions from
  caustic crossing studied in this paper, microlensing observations of
  M31 with Subaru/Hyper Suprime-Cam (HSC) \citep{Niikura:2017zjd},
  EROS/MACHO microlensing \cite{Alcock:2000ph,Tisserand:2006zx}, 
  ultra-faint dwarf galaxies (UFDs) \cite{Brandt:2016aco}, and Planck
  cosmic microwave background observations (Planck)
  \cite{Ali-Haimoud:2016mbv}. \morv{We also show constraints from caustic
  crossing with different assumptions on the transverse velocity
  (factors of 2 and 0.5 different from the fiducial value) by dotted
  lines.} For UFDs and Planck, conservative limits
  are shown by solid lines, whereas more stringent limits are shown by
  dashed lines. \morv{For the Planck constraint, the stringent limit
    assumes the collisional ionization around PBHs whereas the
    conservative limit assumes the photoionization due to the
    PBH radiation. For the UFDs constraint, different constraints
    reflect different assumptions on the dark matter densities and
    initial sizes of star clusters in UFDs.}} 
\label{fig:pbh}
\end{figure}

From this argument, we can derive constraints on the mass $M$ and
abundance $f_{\rm p}$ of compact dark matter. Figure~\ref{fig:pbh}
shows the rough excluded region in the $M$-$f_{\rm p}$ plane from the
observation of MACS~J1149 LS1. As discussed in \cite{Kelly:2017fps},
the very high abundance of $\sim 30~M_\odot$ black holes
\citep{Bird:2016dcv}, which is motivated by recent observations of
gravitational waves \cite{Abbott:2016blz}, is excluded, although more
careful comparisons with simulated microlensing light curves should be
made in order to place more robust constraints.

\morv{As discussed in Section~\ref{sec:uncertainty}, our constraints
  are subject to model parameter uncertainties. While we find that the
uncertainty of the macro model magnification $\mu_{\rm h}$ does not
affect our constraint on $f_{\rm p}$ from the saturation condition,
the uncertainty on the transverse velocity $v$ is expected to have a
large impact on our result via the dependence of the maximum source
star radius on the velocity (equation~\ref{eq:const5}). To evaluate
this, we repeat the analysis presented above with different values of
the transverse velocity by $\pm 0.5$~dex, i.e., factors of 2 and 0.5
different from the original value of $v=500~{\rm km\,s^{-1}}$. The
resulting constraints shown in Fig.~\ref{fig:pbh} indicate that our
results on the compact dark matter abundance are indeed sensitive to
the assumed transverse velocity. We find that the constraint on
$f_{\rm p}$ is weaker for the higher velocity, because
equation~(\ref{eq:const5}) suggests that larger source radii (i.e.,
intrinsically brighter source star) are allowed for the higher
velocity. We find that the very high abundance of $\sim 30~M_\odot$
black holes are still excluded even for the high velocity case, which
is encouraging. In order to draw more robust constraints on $f_{\rm
  p}$, we need to convolve our constraints on the probability
distributions of the transverse velocity as well as other model
parameters, which we leave for future work.}

We expect that we can place tighter constraints on compact dark matter
from long monitoring observations of giant arcs and careful analysis of
observed light curves. This is because point mass lens with different
masses have quite different characteristics of light curves such as
time scales and peak magnifications. Therefore, observations or
absence of light curve peaks with different time scales may be used to
place constraints on the abundance of compact dark matter with
different masses, although we have to take account of the uncertainty
in the velocity for the robust interpretation. As discussed in
\cite{Diego:2017drh}, another clue may be obtained by detailed
observations of light curves before and after the peak.
As mentioned above, in order to obtain robust constraints on compact
dark matter from observations, it is also important to conduct
ray-tracing simulations that include both ICL stars and compact dark
matter, as was partly done in \cite{Diego:2017drh}. Ray-tracing
simulations are helpful to better understand what kind light curves
such compound lens system predict.  

\section{Summary and discussions}
\label{sec:summary}

In this paper, we have adopted a simple analytical lens model that
consists of a point mass lens and a constant convergence and shear
field, which is used to study lensing properties of a point mass lens
embedded in high magnification regions due to the cluster potential.
This model has been used to derive characteristic scales of caustic
crossing events in giant arcs, such as the time scale of light curves
and  maximum magnifications, as a function of the mass of the point
mass lens and the radius of the source star. We have tuned model
parameters to the MACS~J1149 LS1 event to constrain lens and source
properties of this event. We have also computed expected event rates,
and derived additional constraints on the lens and source properties
of  MACS~J1149 LS1.

Our results that are summarized in Figures~\ref{fig:rate} and 
\ref{fig:const_rate} indicate that MACS~J1149 LS1 is fully consistent
with microlensing by ICL stars. The allowed ranges of the lens mass
and source radius are $0.1~M_\odot \lesssim M \lesssim 4\times 10^3M_\odot$
and $40~R_\odot \lesssim R \lesssim 260~R_\odot$, 
respectively. The most plausible radius of the
source star is $R\approx 180~R_\odot$ (luminosity $L\approx
6\times 10^5~L_\odot$), which is consistent with a blue supergiant. 
In this case, the source star should have been magnified by a factor
of $\approx 4300$ at the peak. Our results suggest that the allowed
ranges of the lens mass and source radius are relative narrow, which
limit the possibility of explaining MACS~J1149 LS1 by exotic
dark matter models.

We have discussed the possibility of constraining compact dark matter
in the presence of ICL stars. Using the saturation argument, we have shown
that compact dark matter models with high fractional matter densities
($f_{\rm p}\gtrsim 0.1$) for a wide mass range of $10^{-5}M_\odot
\lesssim M\lesssim 10^2M_\odot$ are inconsistent with the observation
of MACS~J1149 LS1 because such models predict too low magnifications
at the position of MACS~J1149 LS1. We note that this constraint from
the saturation condition should be applicable to the total
compact dark matter fraction for models with extended mass functions
\cite{Carr:2017jsz}. We expect that we can place tighter
constraints on the abundance and mass of compact dark matter by
careful analysis of observed light curves as well as more observations
of caustic crossing events.

In this paper, we have assumed a single star as a source. As discussed
in \cite{Kelly:2017fps}, there is a possibility that the source is in
fact a binary star, based on multiple peaks in the light curve. Even
for a binary star system, our results are broadly applicable to
individual stars that constitute the binary system.

There are several additional caveats in our analysis.
\morv{As discussed in the paper, our constraints sensitively depends
  on the assumption on the the velocity $v$ as well as the macro model
  magnification.  To draw more robust conclusion we have to take
  account of the distributions of the velocity and the macro model
  magnification.} We can also consider more realistic star models, such
  as  an improved mass-radius relation of stars beyond the black body
relation (equation~\ref{eq:blackbody}) and a more realistic population
of stars with various temperatures. 

We also did not discuss the ``counterimage'' (``Iapyx'') of MACS~J1149
LS1 presented in \cite{Kelly:2017fps}. The position of the second image
which was separated by $0.26''$ from MACS~J1149 LS1 is consistent with
being the counterimage. \cite{Kelly:2017fps} argued that a point mass
lens with $M\gtrsim 3~M_\odot$ is needed to demagnify the counterimage
for several years.  From equation~(\ref{eq:betaplus}), we can estimate
the timescale of the demagnification as $t_{\rm demag}\approx
2\beta_{{\rm crit},+}/u\approx 14(M/M_\odot)^{1/2}{\rm year}$, which
suggests that indeed a point mass lens with $M\sim 3~M_\odot$ is
capable of demagnifying the counterimage for many years. In
order for this conclusion to hold, the counterimage must be bright
enough to be detected in absence of microlensing. From
equation~(\ref{eq:mmin}), we can estimate the apparent brightness of
the counterimage without microlensing magnification (but is magnified
by the macro lens model) to be $\approx 28.9$~mag for $R=180~R_\odot$,
which is much fainter than the limiting magnitude of the monitoring
observation. However, the source radius is allowed to be as large as
$R=260~R_\odot$ (see Figure~\ref{fig:const_rate}), which suggests that
the counterimage can be as bright as $\approx 28.1$~mag without
microlensing, which can marginally be detected in individual
observations of  MACS~J1149. In observations, while a source with the
magnitude $\approx 28$ was observed in previous images
\cite{Kelly:2017fps}, given the expected fluctuations of light curves
and the limited time coverage of observations it is not clear whether
this really corresponds to the brightness of the source for the macro
model magnification $\mu_{\rm t}\mu_{\rm r}=300$. Therefore, the conclusion
about the demagnification of the counterimage crucially depends on the
intrinsic radius (luminosity) of the source star. There is also a
possibility that this is in fact not a counterimage but a distinct
star magnified by microlensing. While this may be more plausible given
the relatively low probability of caustic crossing events for
individual source stars, simulations in \cite{Kelly:2017fps} indicate
that a single star tends to be responsible for the vast majority of
the detectable microlensing peaks. This, together with the rarity of
blue supergiant stars, prefers the scenario that Icarus and Iapyx
originates from the same star. 

An additional caveat is that substructures can also change the relative
macro model magnifications of Icarus and Iapyx, as noted in
\cite{Kelly:2017fps}. While the standard cold dark matter naturally
predicts such dark halo substructures, compact dark matter with
relatively large masses can produce more fluctuations on the macro
model magnification due to the Poisson fluctuations of the projected
surface mass density as a function of position on the sky. Such
spatial variation of the macro model magnifications should have impact
on our quantitative results, including constraints on compact dark
matter from the saturation argument (Section~\ref{sec:compactdm}). We
leave the exploration of this effect in future work.

To summarize, our analytic examinations have demonstrated that
observations of caustic crossing events in giant arcs have a great
potential to study the nature of dark matter. Our predictions on
characteristic scales and event rates should provide useful guidance
for future monitoring of giant arcs in clusters for obtaining various
constraints from caustic crossing events.

\acknowledgments 
We thank an anonymous referee for useful suggestions.
This work was supported in part by World Premier International
Research Center Initiative (WPI Initiative), MEXT, Japan, and JSPS
KAKENHI Grant Number JP26800093 and JP15H05892.
J.~M.~D. acknowledges the support of projects AYA2015-64508-P
(MINECO/FEDER, UE), AYA2012-39475-C02-01, and the consolider project
CSD2010-00064 funded by the Ministerio de Economia y Competitividad.

\end{document}